\documentclass{optica-article}

\journal{opticajournal}

\articletype{Research Article}

\usepackage{algorithm}
\usepackage{algorithmic}

\usepackage{lineno}

\begin{document}

\title{Far-field diffraction computational imaging based on parameter-robust illumination and direct phase optimization}

\author{An-dong Xiong,\authormark{1,2} Xiao-peng Jin,\authormark{1}  Xv-ri Yao,\authormark{1}and Qing Zhao\authormark{1,*}}

\address{\authormark{1}Center for Quantum Technology Research, School of Physics, Beijing Institute of Technology, Beijing 100081, China\\
\authormark{2}xiongandong1994@outlook.com}

\email{\authormark{*}qzhaoyuping@bit.edu.cn}

\begin{abstract}
Coherent diffraction imaging (CDI) is a promising imaging technique revealing most of the information from diffraction measurements. An ideal CDI should reconstruct complex-valued object from a single-shot far-field diffraction without any priori information about the target. To realize the ideal CDI, we propose a class of parameter-robust illumination pattern. A direct phase optimizing algorithm is also raised here to improve the performance of phase retrieval in strong noise. Experimental result demonstrates the efficiency of our scheme in practical noisy measurement for complex-valued target. 
\end{abstract}

\section{Introduction}
The Fraunhofer diffraction or far-field diffraction can be easily achieved in most scenarios, including visible light, X -ray and electron\cite{shechtman2015phase}. CDI utilizing far-field diffraction is widely applied in observing structure of microorganism\cite{tian2014multiplexed,gallagher2014macromolecular,song2008quantitative,rodriguez2013oversampling}, molecule\cite{loh2012fractal,chapman2011femtosecond} and material\cite{miao1999extending,pfeiffer2018x,obinson2001reconstruction}. The mathematical form of the diffraction field is concisely the Fourier transformation of the light field at the object plane. However, because the phase part of the light field is hardly directly measurable\cite{shechtman2015phase}, it needs to be retrieved for the Fourier inversion. In practice, as a multi -solution problem, phase retrieval is usually conducted in two ways. One is multiple measurements: several masks cast on the target\cite{candes2015phasewf,tian2014multiplexed,yu2017efficient} or overlap scanning such as ptychography\cite{rodenburg2008ptychography,holler2014x,pfeiffer2018x,gardner2017subwavelength} to provide more information. The other way is to use certain generals of the object field in specific application scenarios, such as real-valued condition\cite{gauthier2010single,loh2012fractal,duarte2019computed,Mansi2019Phase}, sparsity in certain transform\cite{donoho2006compressed,ohlsson2012cprl,shechtman2014gespar,sidorenko2015sparsity} and deep learning methods\cite{barbastathis2019use,chen2022physics}. Multiple measurement needs to ensure that the object will not change between the measurements, which limits dynamic imaging and high energy imaging\cite{sayre1995x}. Utilization of priori information surely cannot faithfully reconstruct the object that violate these characteristics, for example, a random complex-valued field. If the far-field diffraction measurement and computational reconstruction is taken as a black box, it is hoped to be as simple as a mirror or lens: simply reflects any original field without multiple measurements or priori information. Efforts have been made for the ideal black box and achieve good simulation result\cite{fienup1978reconstruction，fienup1982phase, miao1998phase}. But when applied in practice, extra restriction seems indispensable. Intensity object is still more welcome in experiments\cite{miao1999extending,gauthier2010single,barty2008ultrafast}.

The phase retrieval of single-shot far-field diffraction without prior information is inherently a problem with multiple equivalent solutions\cite{guizar2012understanding}. Some equivalent solutions cannot be excluded by simple oversampling, including:\begin{itemize}
\item overall phase shift;
\item translation paired with phase modulation that matches the moving distance;
\item inversion of complex conjugate;
\item combinations of the operations above.
\end{itemize}
In the presence of higher noise, the resulting final phase retrieval solution tends to be degraded by superimposed mixture of the equivalent solutions.

In order to suppress the degradation, there have been some good simulation works using special illumination pattern to destroy certain symmetry\cite{fienup1987reconstruction,fienup2006lensless}. These methods can also be classified into adding general information about the object field. Because the special illumination is usually built-in the set-up, the information remains unchanged, providing more feasibilty for different samples. But it appears difficult to achieve satisfying result in practical applications. Special field restriction is still more viable for intensity sample\cite{gauthier2010single,zhang2016phase} . The perfect pixel mapping and size parameters of the pattern and object can be easily obtained in the simulation calculation. However, due to the optical path distortion, aberration, fluctuation of refractive index, photosensitive array spacing and other practical problems in experiment, the parameter and pixel mapping used in the reconstruction algorithm are hardly as precise as those in simulation. These errors can greatly degrade the retrieval result under high noise as shown in Fig. 2(b). Here, we propose a class of parametric-robust apertures and a direct phase optimization algorithm for the corresponding special-shaped apertures, together with a measurement correction algorithm.

\section{Illumination pattern}

For phase retrieval without priori information, all we can rely on is that the intensity of the light field outside the passable region is zero. In the optimization procedure, if the imagined aperture is smaller than the true one, the actual light intensity in the assumed zero zone will not be zero, which will mislead the algorithm to solve an impossible problem. Therefore, it is necessary to ensure that the theoretical aperture can cover the actual aperture. Due to parameter errors, the ideal aperture needs to be slightly enlarged. . We choose linear magnification operation to discuss hereinafter. The reason is:
1.Operations such as dilation and convolution will lead to the emergence of translation equivalent solutions;
2.Errors like boundary error on Fourier plane will lead to linear zoom on image plane;
3.The linear magnification operation retains most of the symmetry embodied in the measurement.

The linear magnification operation is: for a enlarge center point $O$ and any point $A$ in the space, the corresponding new point $A '$ needs to satisfy $ \vec{OA '} = a*\vec{OA} \ \ (a>1)$. We require that for any $a>1$, the enlarged space covers the original one, there is $\{A\} \subset \{A'\}$. Let the boundaries be $\{L_i\}$ for the aperture space. These $\{L_i\}$ are Jordan curves. Jordan curve theorem will be used a lot below: every continuous path connecting ‘outside’ and ‘inside’ points intersects with the curve somewhere. 

For $O$ outside a certain $L_i$, if $O$ is inside $L_i’$, because $O=O’$, $O’$ is inside $L_i’$; $O$ must be inside $L_i$; but $O$ is outside $L_i$; contradiction occurs. Therefore, If $O$ is outside $L_i$, $O$ must be outside $L_i’$.

If $O$ is outside a certain $L_i$, $Q$ is a point inside $L_i$, the nearest intersection of line segment $OQ$ and $L_i$ to $O$ is $A$(Fig. 1(b)). If $A$ is on or inside $L_i’$, because $O$ must be outside $L_i’$, $OA$ and $L_i’$ should intersect at some point $G’$. $OA>OG’>OG$ ($OG$ and $OG'$ must be in the same direction, outerwise $O$ is inside $L_i’$(Fig. 1(c))), and $G$ is on $L_i$, so $A$ is not the nearest intersection, which is a contradiction. So $A$ is outside $L_i’$, which means $L_i$ cannot be completely surrounded by $L_i’$. For multi-connected graphs, there may be an a large enough that $L_j '$ can surround $L_i$. But we need $a$ to be any number more than 1. For a multiply connected graph, for any $O$, there is always some $L_i$ that $O$ is outside $L_i$. Therefore, only a simply connected graph can meet our needs.

And for this simply connected graph, we require that: there is at least one $O$ inside $L$ that any ray from $O$ intersects $L$ at only one point. We call $O$ a simple encircled center. For this kind of graph, suppose that $C$ is in $\{A\}$ but not in $\{A '\}$, let $B_1  '$ and $B_2 '$ be the intersection of line $OC$ and $L '$ , $OB_1 '$ and $OC$ are in the same direction. If $B_1 $ does not belong to $\{A’\}$, $B_1 O$ intersects $L’$ on $G’$, $OG<OG’<OB_1$, $G \neq B_1 $ ,contradicting simple encircling. Hence $B_1 \in \{A '\}$. That means, there is at least one intersection $D '$ on $B_1 C$ and $L '$ . The point $D$ corresponding to $D'$ is the intersection of line $OC$ and $L$ , contradicting simple encircling (Fig. 1(a)). Therefore, for a simply connected graph that meets the above conditions, its linear magnification of $O$ according to any $a>1$ can completely cover the original graph. Incidentally, not every simply connected graph has a simple encircled center, for example, a slim $\Omega$.

\begin{figure}[H]
\centering\includegraphics[width=1.00\textwidth]{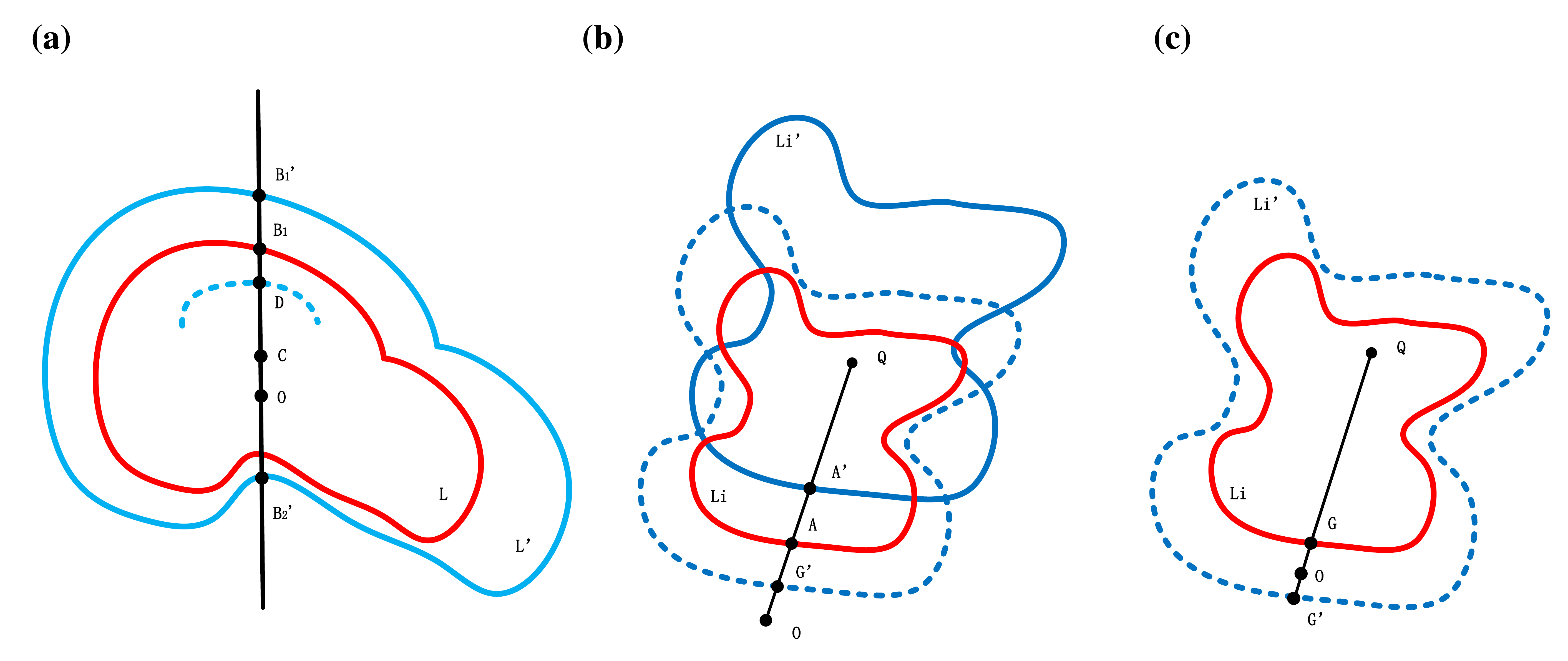}
\caption{Schematic for the proof of the features of linear magnification. The red curve is the original curve while the blue curve is the magnified one. the dotted curve is a fake linear magnified curve for reductio ad absurdum, which is actually magnified according to some different $O$}
\end{figure}

In order to satisfy the requirements above, we can generate several sectors centre on point $O$ to generate the desired aperture shape. The radius, angle and number of these sectors are not limited, which means shapes like rectangle and triangle can also be generated.

In addition to the requirement that linear magnification fully covers the original one. The aperture shape had better have the following properties:\begin{itemize}
\item It has a small self-convolution maximum value (corresponding to the maximum overlap with the inversion after translation);
\item When enlarged, it limits the possible translation of the original graphics in it;
\item Continuous area in it is as large as possible so that the final reconstructed image is more identifiable.
\end{itemize}
Based on the conditions described above, we choose the aperture of the equal-radius three-sector structure as an example below. The maximum value of self-convolution for three-sectors is 0.484 while that for five-sectors only decreases to 0.477. The maximum self-convolution value for more number of odd-numbered sectors even increases.

Fig. 1 shows the noiseless performance of multiple connected non-centrosymmetry aperture and simple connected three-sectors aperture. When the assumed aperture is exactly the actual one, both of them works well in attaining the original image. However, when the assumed aperture is slightly enlarged, the result is dramatically degraded in the multiple connected case, because the enlarged pattern cannot cover the original one. While three-sectors after enlarged contains the original graph, leading to robustness towards size error.

\begin{figure}[H]
\centering\includegraphics[width=1.00\textwidth]{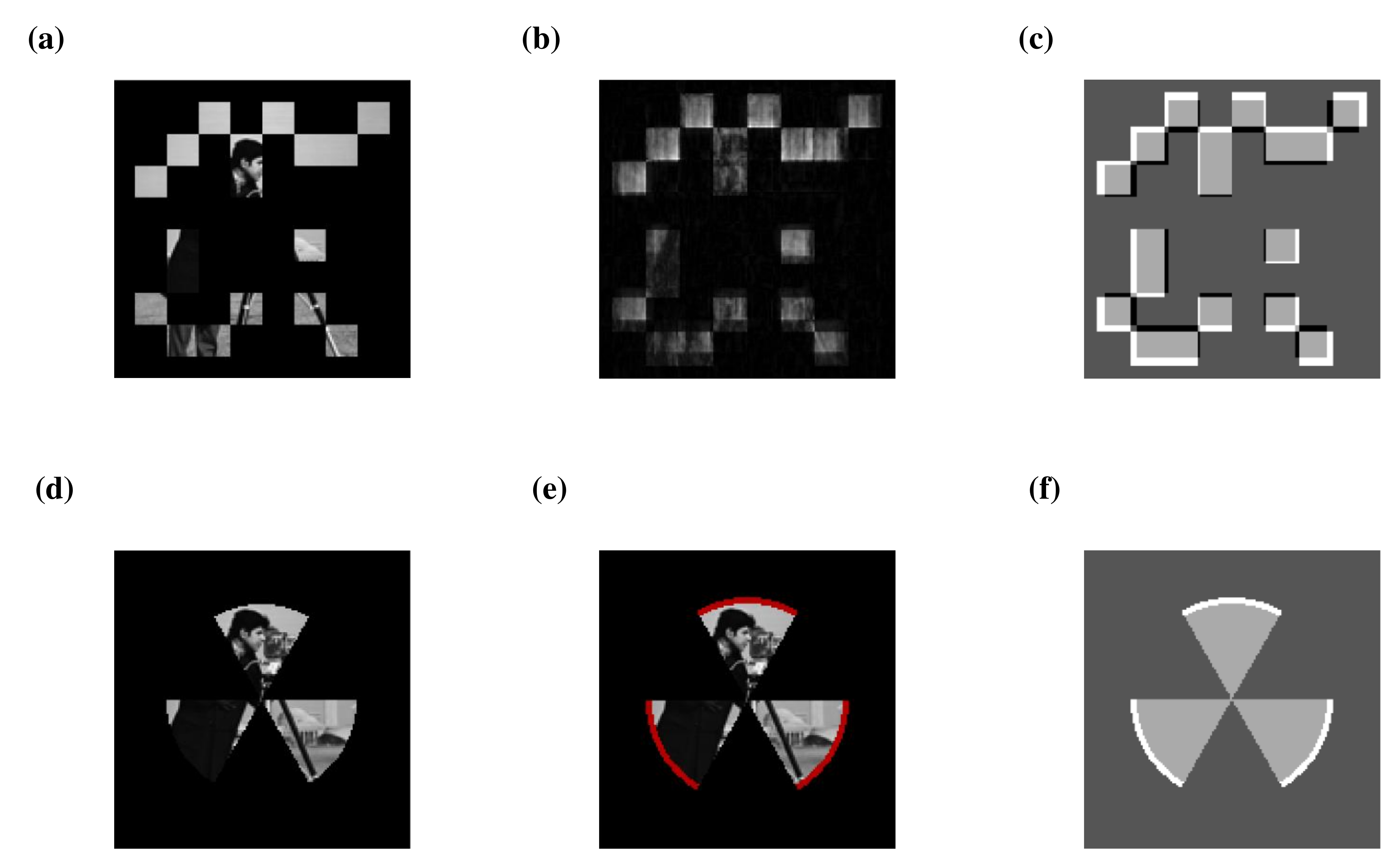}
\caption{Simulation reconstruction of OPM from noise-free measurement of the 'cameraman'. (a),(b) take a random distributed multiply connected as he illumination pattern, while (d),(e) use three-sectors. (a),(d) are the reconstrunctions when precise-sized pattern is given. (b),(e) are the reconstructions when $a = 17/16$. (d) and (e) are acually the same, the red part is the enlarged area of the given pattern. (c),(f) show the overlap after magnification. The black pixels in (c) are those not covered by the magnified pattern.}
\end{figure}

\section{Algorithm optimizing phase and measurement}
In this paper, no prior information about the target object is given. The only information is the intensity of the light field outside the aperture is zero or as low as possible. Without the assistance of other information, traditional phase retrieval algorithms such as alternative-projection and gradient algorithms are difficult to stably converge to a better solution for complex-valued object under high noise\cite{guizar2012understanding,latychevskaia2018iterative}. Let the light field at the wave vector $k$ in the Fourier plane be ${m_k}{e^{i{\theta _k}}}$. Many algorithm works take ${e^{i{\theta _k}}}$ as a whole to optimize the loss value, so that many linear optimization methods can be utilized\cite{goldstein2017convex,candes2013phaselift,candes2015phase}. But in the end, The quadratic condition that the modulus of ${e^{i{\theta _k}}}$ is 1 has to be used, which is the principal trouble. In order to maximize the use of aperture information, we take $\theta _k$ as the optimization object here.

The definitions that will be used later are collected here to be checked.
\begin{equation}
s = {\left[ {{e^{i\theta {k_1}}}, \ldots ,{e^{i\theta {k_N}}}} \right]^T}
\end{equation}

\begin{equation}M = diag\left( {{m_1}, \ldots ,{m_N}} \right)\end{equation}
\begin{equation}{F_{\vec k\vec j}} = {e^{ - i\vec j\vec k}}\end{equation}
\begin{equation}{f_{\vec j\vec k}} = {e^{i\vec j\vec k}}\end{equation}
\begin{equation}\widehat F{\left[ x \right]_{\vec k}} = \sum\limits_{\vec j} {{F_{\vec k\vec j}}{x_{\vec j}}} \end{equation}
\begin{equation}\widehat f{\left[ X \right]_{\vec j}} = \sum\limits_{\vec k} {{f_{\vec j\vec k}}{X_{\vec k}}} \end{equation}
\begin{equation}{g_{jk}} = {m_k}{e^{ijk}}\end{equation}
\begin{equation}{v_j} = \sum\limits_k {{g_{jk}}{e^{i{\theta _k}}}} \end{equation}
\begin{equation}\gamma_k  = \sum\limits_{j \in Z} {\left( {{g_{jk}}\sum\limits_{k' \ne k} {{g_{jk'}}^*{e^{ - i{\theta _k}^\prime }}} } \right)} \end{equation}
\begin{equation}D_j^{\left( k \right)} = \sum\limits_{k' \ne k} {{g_{jk'}}} {e^{i{\theta _k}^\prime }}\end{equation}
\begin{equation}E_1^k = diag(\underbrace {1, \ldots ,1}_{k - 1},0,\underbrace {1, \ldots 1}_{N - k})\end{equation}
\begin{equation}{\xi _2}(j') = \left\{ {\begin{array}{*{20}{c}}
{1{\rm{   ,  }}j' \in Z}\\
{0{\rm{   ,  }}j' \notin Z}
\end{array}} \right.\end{equation}
\begin{equation}{E_2} = diag({\xi _2})\end{equation}

Let $Z$ denote the set of points unilluminated in object plane.The loss value for the aperture information is expressed as
\begin{equation}L = \sum\limits_{j \in Z} {{v_j}{v_j}^*} \end{equation}
 and its partial derivative with respect to $\theta _k$ is:
\begin{equation}
\begin{aligned}
\frac{{\partial L}}{{\partial {\theta _k}}} =  & \sum\limits_{j \in Z} {\left( {\frac{{\partial {v_j}}}{{\partial {\theta _k}}} \cdot {v_j}^* + \frac{{\partial {v_j}^*}}{{\partial {\theta _k}}} \cdot {v_j}} \right)} \\
 =  & \sum\limits_{j \in Z} {\left( {i{g_{jk}}{e^{i{\theta _k}}} \cdot {v_j}^* - i{g_{jk}}^*{e^{ - i{\theta _k}}} \cdot {v_j}} \right)} \\
 =  & \sum\limits_{j \in Z} {2real\left( {i{e^{i{\theta _k}}}{g_{jk}} \cdot {v_j}^*} \right)} \\
 =  & \sum\limits_{j \in Z} {2real\left( {i{e^{i{\theta _k}}}{g_{jk}} \cdot \sum\limits_{k'} {{g_{jk'}}^*{e^{ - i{\theta _k}^\prime }}} } \right)} \\
 =  & \sum\limits_{j \in Z} {2real\left( {i{e^{i{\theta _k}}}\left( {{g_{jk}} \cdot \sum\limits_{k' \ne k} {{g_{jk'}}^*{e^{ - i{\theta _k}^\prime }}}  + {g_{jk}}{g_{jk}}^*{e^{ - i{\theta _k}}}} \right)} \right)} \\
 =  & \sum\limits_{j \in Z} {2real\left( {i{e^{i{\theta _k}}}\left( {{g_{jk}} \cdot \sum\limits_{k' \ne k} {{g_{jk'}}^*{e^{ - i{\theta _k}^\prime }}} } \right) + {{i\left| {{m_k}} \right|}^2}} \right)} \\
 =  & \sum\limits_{j \in Z} {2real\left( {i{e^{i{\theta _k}}}\left( {{g_{jk}} \cdot \sum\limits_{k' \ne k} {{g_{jk'}}^*{e^{ - i{\theta _k}^\prime }}} } \right)} \right)} \\
 =  & 2real\left( {i{e^{i{\theta _k}}}\gamma_k } \right)
\end{aligned}
\end{equation}

Since $\gamma _k$ does not contain $\theta _k$ (Eq. (9)), 
\begin{equation}\frac{{\partial \gamma_k }}{{\partial {\theta _k}}} = 0\end{equation}
The value of $\theta _k$ at the extreme point can be directly solved by the condition that the partial derivative is 0.
Let
\begin{equation}\gamma_k  = \rho {e^{i\alpha }}\end{equation}
At the extreme point
\begin{equation}\frac{{\partial L}}{{\partial {\theta _k}}} = i{e^{i{\theta _k}}}\rho {e^{i\alpha }} - i{e^{ - i{\theta _k}}}\rho {e^{ - i\alpha }} =  - 2\rho \sin \left( {{\theta _k} + \alpha } \right) = 0\end{equation}

Because
\begin{equation}\frac{{{\partial ^2}L}}{{\partial {\theta _k}^2}} =  - 2\rho \cos \left( {{\theta _k} + \alpha } \right)\end{equation}

At the local minimum, we have
\begin{equation}{\theta _k} =  - \alpha  + \pi \end{equation}
\begin{equation}{e^{i{\theta _k}}} =  - {\left( {{e^{i\alpha }}} \right)^*}\end{equation}

An equation for $\theta _k$ at the local minimum is already achieved here. However, if  $\gamma _k$  is calculated separately for each wave vector pixel in the Fourier measurement, the overall phase retrieval may require millions of Fourier transforms, so we need to go further to reduce the calculation amount.
\begin{equation}
\begin{aligned}
{\gamma _k} =  & \sum\limits_{j \in Z} {\left( {{g_{jk}}D{{_j^{\left( k \right)}}^*}} \right)} \\
 =  & \sum\limits_j {{m_k}{e^{ijk}}{\xi _2}(j)D{{_j^{\left( k \right)}}^*}} \\
 =  & {m_k}{\left( {\sum\limits_j {{e^{ - ijk}}{\xi _2}(j)D_j^{\left( k \right)}} } \right)^*}
\end{aligned}
\end{equation}
Let
\begin{equation}\Gamma  = {\left[ {{\gamma _1}, \ldots ,{\gamma _N}} \right]^T}\end{equation}
\begin{equation}
\begin{aligned}
\Gamma  =  & {\left( {M\hat F{E_2}\hat fM{E_1}s} \right)^*}\\
 =  & M\left( {\hat F{E_2}\hat fMs - \sum\limits_{j \in Z} {Ms} } \right)^*\\
 =  & M\left( {N\hat F'{E_2}\hat f'Ms - \sum\limits_{j \in Z} {Ms} } \right)^*
\end{aligned}
\end{equation}

Here $\hat F'$ and $\hat f'$ are two demensional FFT (Fast Fourie Transform) and iFFT (inverse Fast Fourie Transform).

Since it is a multivariate problem, the solution obtained after each batch is the extreme point of the previous batch of phases, so it still takes several iterations to converge to the extreme point. For each pixel, if only one $\theta_k$ is changed at a time, the new overall loss value obtained after each calculation will only be less than or equal to the value of the previous batch. Therefore we can have confidence in the convergence to this problem. Of course, considering the actual calculation efficiency, it is not necessary to be too cautious here, and the phase value can be updated more comprehensively.

In high noise, the extreme point is not necessarily the global lowest point. Since we are confident in the convergence, we can get away from local minima by resetting the phase values of a large percentage of random locations to random phases when the solution process stalls.

\begin{algorithm}[H]
\caption{Optimizing phase for measurement (OPM)}
\label{alg:OPM}
\begin{algorithmic}
\STATE \textbf{Input:} $M$, $E_2$, $S^0$
\STATE $s^{best} \leftarrow s^0$
\FOR {$t=1$ to $T$}
\STATE $\Gamma ^t \leftarrow  \left( {N\hat F'{E_2}\hat f'Ms^t - \sum\limits_{j \in Z} {Ms^t} } \right)^*$
\STATE  $s^{t} \leftarrow \begin{cases}
- {(\frac{{{\Gamma ^t}}}{{\left| {{\Gamma ^t}} \right| + \mu }})^*}& \text{ 90\% location } \\
s^{t-1}& \text{ 10\% location } 
\end{cases}$

\STATE\COMMENT {The $\mu$ value not only prevents dividing by zero, but also has the effect of a noise filter.}
\IF {$L(M,{s^t},{E_2}) < L(M,{s^{best}},{E_2})$}
\STATE  $s^{best} \leftarrow s^t$
\ENDIF
\IF {$L(M,{s^{t-1}},{E_2}) - L(M,{s^{t}},{E_2}) < threshold$}
\STATE  $counter \leftarrow counter+1$
\ENDIF
\IF {$counter>counterthreshold$}
\STATE  $s^{temp} \leftarrow \begin{cases}
s^t& \text{ 70\% chance } \\
s^{best}& \text{ 30\% chance } 
\end{cases}$
\STATE $s^{t} \leftarrow randomize(s^t,90\%)$
\ENDIF
\ENDFOR
\STATE \textbf{Output:} $S^{best}$
\end{algorithmic}
\end{algorithm}

On the other hand, we can also optimize the measurement to some extent in a similar way.
(loss to mk)
\begin{equation}
\begin{aligned}
\frac{{\partial L}}{{\partial {m_k}}} =  & {e^{i{\theta _k}}}\sum\limits_{j \in Z} {{e^{ijk}}{v_j}^*}  + {e^{ - i{\theta _k}}}\sum\limits_{j \in Z} {{e^{ - ijk}}{v_j}} \\
 =  & 2real\left( {{e^{ - i{\theta _k}}}\sum\limits_{j \in Z} {{e^{ - ijk}}\sum\limits_{k'} {{e^{ijk'}}{m_k}^\prime {e^{i{\theta _k}^\prime }}} } } \right)\\
 =  & 2real\left( {{e^{ - i{\theta _k}}}\widehat F{E_2}\widehat fMs} \right)
\end{aligned}
\end{equation}
The measurement value at extreme point under the current phase can be directly obtained utilizing the linear expression of $M$. However, the direct calculation will lead to a wrong measurement when a wrong phase distribution is given. Error in phase cause error in measurement; error in measurement cause error in phase, causing error to be fixed in subsequent iterations. So only limited optimization can be implemented based on the existing gradient. At the same time, because loss value corresponding to all 0 measurement is 0, there is always a tendency to reduce the overall measurement intensity. Hence, change on sum of the measurement values must be restricted. So we adopt the following update strategy here to assign the change value according to the gradient.

\begin{equation}
\Delta {m_k} =  - q \cdot \left( {\frac{{\frac{{\partial L}}{{\partial {m_k}}}}}{{\sum\limits_k {\frac{{\partial L}}{{\partial {m_k}}}} }} - \frac{1}{N}} \right)
\end{equation}
Where $q$ is the the variable controlling the overall gradient. In addition, we limit the maximum change during the iteration process to make measurement as faithful as possible to the original one.

\begin{algorithm}[H]
\caption{Optimizing measurement for phase (OMP)}
\label{alg:OMP}
\begin{algorithmic}
\STATE \textbf{Input:} $M^0$, $E_2$, $S$, $amc$, $rmc$
\STATE ${M_{\min }} \leftarrow \max (\min ((1 - rmc) \bullet {M^0},({M^0} - amc)),0)$
\STATE ${M_{\max }} \leftarrow \max ((1 + rmc) \bullet {M^0},({M^0} - amc))$
\FOR {$t=1$ to $T$}
\STATE $\frac{{\partial L}}{{\partial {m_k}}} \leftarrow 2real\left( {{e^{ - i{\theta _k}}}\widehat F{E_2}\widehat fMs} \right)$
\STATE $\Delta {m_k} \leftarrow  - q \cdot \left( {\frac{{\frac{{\partial L}}{{\partial {m_k}}}}}{{\sum\limits_k {\frac{{\partial L}}{{\partial {m_k}}}} }} - \frac{1}{N}} \right)$
\STATE $ M^{t'} \leftarrow M^{t-1}+\Delta M$
\STATE ${M^t} \leftarrow \min ({M_{\max }},\max ({M^{t'}},{M_{\min }}))$
\ENDFOR
\STATE \textbf{Output:} $M^{t}$
\end{algorithmic}
\end{algorithm}

In Fourier measurements, there is a significant difference between the intensity in the low -frequency area and the high -frequency area. The traditional Signal-Noise Ratio (SNR) can hardly reflect the impact of noise on the object plane. We define Inverse Fourier Signal-Noise Ratio (IFSNR):
\begin{equation}
IFSNR = 10{\log _{10}}(\frac{{\sum\limits_{\vec j}^N {{{\left| {\sum\limits_{\vec k}^N {{e^{i\vec j\vec k}}m_{\vec k}^o{e^{i\theta _{\vec k}^s}}} } \right|}^2}} }}{{\sum\limits_{\vec j}^N {{{\left| {\sum\limits_{\vec k}^N {{e^{i\vec j\vec k}}m_{\vec k}^r{e^{i\theta _{\vec k}^s}}}  - \sum\limits_{\vec k}^N {{e^{i\vec j\vec k}}m_{\vec k}^o{e^{i\theta _{\vec k}^s}}} } \right|}^2}} }})
\end{equation}
Here ${m^o}$ is the noiseless measurement; ${m^r}$ is the noisy measurement; $\theta ^s$ is the supplementary phase. The supplementary phase can be selected as the original Fourier phase, random phase, or the phase of the retrieval. In the simulation, the supplementary is the phase of the noise -free original Fourier transform .

Fig. 3 shows the performance of OPM and OMP under high noise measurement. The procedure in this simulation is OPM $\rightarrow$ OMP $\rightarrow$ OPM $\rightarrow$ OMP. As shown in the Fig. 3(a)(c), even though the high -frequency part has been severely damaged, OPM can still obtain a relatively good result. While OMP also reconstructs many of the destroyed characteristics in high -frequency area.

\begin{figure}[H]
\centering\includegraphics[width=1.0\textwidth]{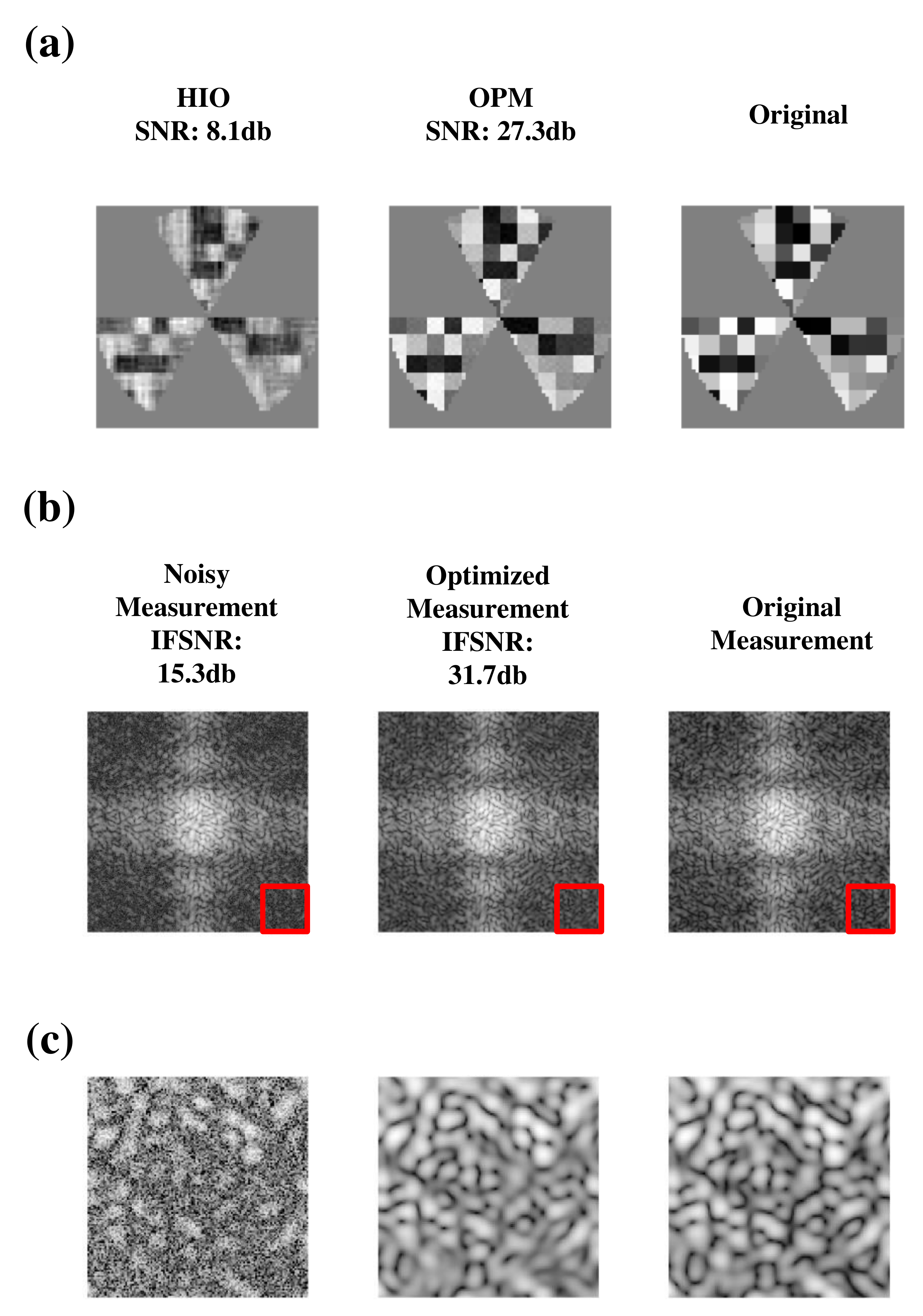}
\caption{Simuation reconstruction from noisy measurement of a random complex-valued matrix. (a) Rconstructed and original imagnary part at object plane. (b) The measurement before and after optimizaion and the original one for comparison, FFTshifted and logarithmically shown. (c) Enlarged red part in (b)}
\end{figure}

\section{Experiment}
\begin{figure}[H]
\centering\includegraphics[width=0.9\textwidth]{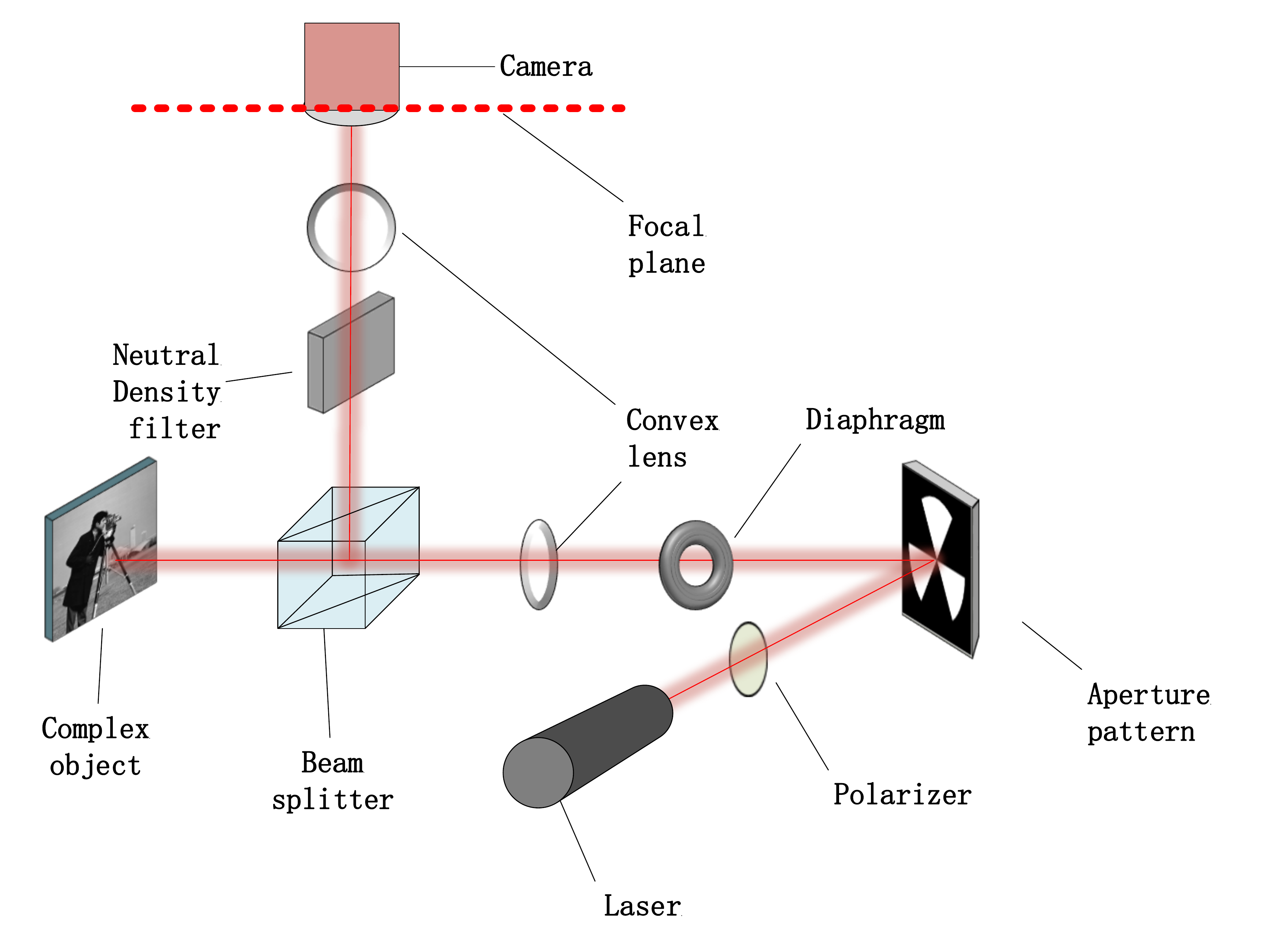}
\caption{Far-field diffraction set-up.}
\end{figure}

The simplified optical path of the set-up is shown in Fig. 4.

The light source is a polarized He-Ne laser light source. The aperture is realized by a pattern loaded on the digital micromirror device (DMD). The aperture is imaged on the spatial light modulator (SLM) through a convex lens. Complex-valued object is a pattern on the SLM. The far-field diffraction pattern is measured with a 14-bit cmos at the focal plane of the lens after SLM reflection. The exposure time is 20ms.

The measurement on CMOS contains some structured noise, hence we conduct OMP first with strong restriction to decrease the noise out of aperture Fourier's transformation. The supplementary phase here is the mean of the Fourier phases of several random patterns with different granularity.

Distortions on the light field that actually projected on the target cannot be perfectly described by linear magnification. Therefore, we first take an assumed aperture with larger angle and radius for reconstruction, and obtain a new aperture based on the reconstruction. Several steps of erosion and dilation\cite{marchesini2003x} are conducted to eliminate scattered pixels to meet the requirements of simple connection.

The coordinate of the Fourier transform center is hardly an just integer on CMOS, result in ripple phase distribution correlated with the offset. Meanwhile, due to surface unevenness, aberration and oblique light path, there is a base phase pattern on the target. These problems are solved by subtracting the reconstructed phase of a full 1 target, in other word, a mirror.

The experiment result is demonstrated in Fig. 5. The OPM faithfully reconstruct the phase morphology of the complex-valued object, showing more details than Hybrid Input Output (HIO). The pillars are also better distinguishable as shown in Fig. 5(e). 
\begin{figure}[H]
\centering\includegraphics[width=1.00\textwidth]{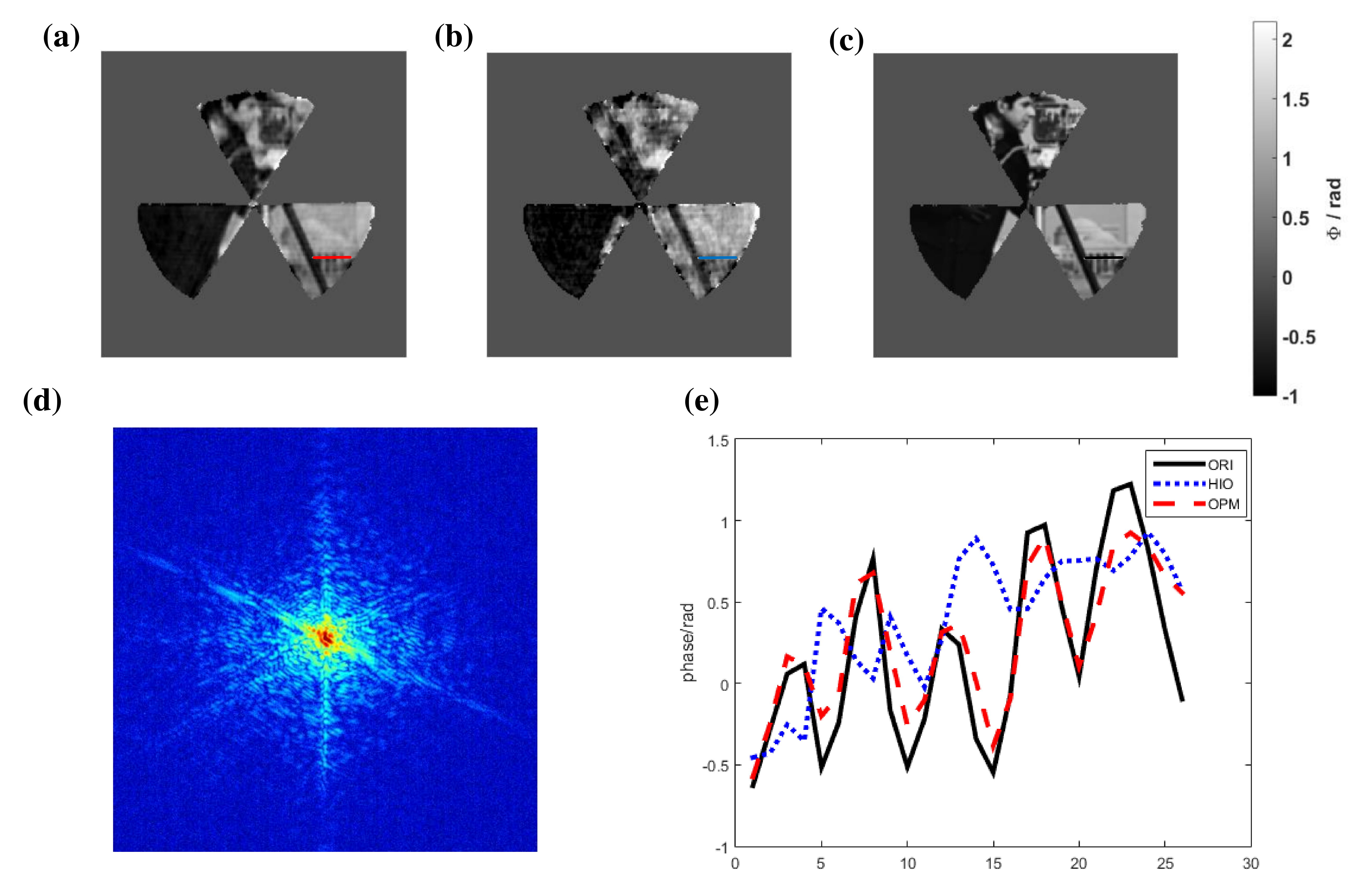}
\caption{Reconstruction from experiment measurement of a complex-valued 'cameraman' cast on the SLM. (a) Reconstrution of OPM; (b) reconstrution of HIO;(c) 'original' object for comparison (due to the unevenness of the actual magnification of the aperture and object, the location relationship between them is roughly achieved by the reconstruction). (a),(b) are phase unwrapped\cite{maier2015robust} and monolithic shifted. (d) Measurement achieved at the CMOS, logarithmically shown (e) The phase value at the bars with different colors in (a),(b),(c)}
\end{figure}

\section{Conclusion}
In summary, we demonstrate that for the parametrically robustness of special illumination method in CDI, the aperture pattern should be simple connected and have a simple encircled center. Three-sector scheme is raised as an appropriate choice. Meanwhile, we propose an algorithm with Fourier phase as the direct optimization object and a coupling measurement correction algorithm. These methods make it possible to achieve phase retrieval with good convergence for complex-valued objects in high-noise measurements. Finally, the experiment result demonstrates the feasibility and efficacy of our scheme and algorithm.

Our aperture scheme only needs blocking and transmitting, hence it can be easily achieved in various wavelengths. Because no information except for the aperture is required in our algorithm, the target is not necessary to meet certain nature such as real-valued and sparsity. Therefore, the algorithm has good compatibility with various application scenarios. A single-shot imaging without multi -frame modulation enables it to be applied to high-energy destructive imaging and imaging of moving objects. The better imaging performance of complex-valued object is of great use in 3D imaging and non-staining biological imaging.

In the algorithm, we use random resetting method to avoid local optimum. Simulated annealing without cooling down consumes a lot calculation time. Better optimization is anticipated here. Subsequent jobs might take deep learning and other methods to fill up the overexposure. Combination of prior knowledge should bring better image quality in specific practical scenarios.

\begin{backmatter}
\bmsection{Funding}
funding

\bmsection{Acknowledgments}
We thank Rong-ping Deng, Wen-kai Yu, Zu-xie Hu and Biao-xv Peng for their help in writing and theory assistance.

\bmsection{Disclosures}
The authors declare no conflicts of interest.

\bmsection{Data Availability}
The data that support the plots within this paper and other findings of this study are available from the corresponding author upon reasonable request.

\end{backmatter}

\nocite{*}
\bibliography{myref}

\end{document}